\documentclass[runningheads]{llncs}
\usepackage[acronym]{glossaries}
\usepackage[T1]{fontenc}
\usepackage{float}
\usepackage{cite}
\usepackage{svg}
\usepackage{graphicx}
\usepackage{hyperref}
\usepackage{tabularx}
\usepackage{subfig}
\usepackage{multicol}
\usepackage{booktabs}
\usepackage{svg-extract}

\DeclareUnicodeCharacter{221E}{\ensuremath{\infty}}

\newacronym{CPES}{CPES}{Cyber-Physical Energy Systems}
\newacronym{MAS}{MAS}{Multi-Agent Systems}
\newacronym{DER}{DER}{Distributed Energy Resources}
\newacronym{PV}{PV}{photovoltaic}

\begin{document}
%\printnoidxglossaries
%
\title{Architectures for Robust Self-Organizing Energy Systems under Information and Control Constraints}
\titlerunning{Architectures for Robust Self-Organizing Energy Systems}
\author{Emilie Frost\inst{1,2}\orcidID{0000-0003-4791-2333} \and
Astrid Nieße\inst{1,2}\orcidID{0000-0003-1881-9172}}
\authorrunning{Frost et al.}
\institute{Department of Computing Science, Carl von Ossietzky Universität Oldenburg, Ammerländer Heerstraße 114-118, Oldenburg, Germany\\
 \and
Distributed Artificial Intelligence, OFFIS - Institute for Information Technology, Escherweg 2, Oldenburg, Germany\\
\email{\{emilie.frost, astrid.niesse\}@uni-oldenburg.de}
} 
\maketitle              % typeset the header of the contribution
\begin{abstract}
Applying the concept of controlled self-organization in agent-based Cyber-Physical Energy Systems (CPES) is a promising approach to ensure system robustness. By introducing an observer/controller architecture to the system, this concept allows for self-organization while still enabling intervention when disturbances occur. Thus, it is possible to respond to effects of cyber attacks, a major threat to current energy systems. However, when implementing an observer to monitor the system and a controller to execute actions for controlled self-organization in CPES, it is essential to take into account restrictions on information and actions resulting from the privacy of local distributed energy resources, regulatory constraints, and data exchange requirements. For this reason, this paper presents architecture variants for the observer and controller that take into account restrictions on access to information and limited actions. In addition, it evaluates possible controller actions in various architectures. The results underscore the importance of considering observer/controller architectures when designing agent-based systems to ensure their robustness for real-world applications.
\end{abstract}
\section{Introduction}\label{sec:intro}
Current energy systems are evolving into \gls{CPES}\footnote{\gls{CPES} as cyber-physical energy systems is a term that should stress the cyber-physical character of current energy systems. Correctly spoken, \gls{CPES} are cyber-physical systems of systems (CPESoS). For reasons of brevity, this is shortened to \gls{CPES}.}, fully integrating physical energy system infrastructures with digital technologies. This brings new vulnerabilities, particularly in the form of cyber attacks that pose significant threats to the stability and security of these systems \cite{peng2019survey, cao2020survey}.
Due to the distributed nature of \gls{CPES}, \gls{MAS} are well suited for various use cases, as system management and control \cite{ren2021multi}.
To ensure robustness against potential threats, these systems can incorporate principles of controlled self-organization \cite{chaabanrobustmas}.
Controlled self-organization comes from the field of Organic Computing and allows for self-organization of the system while still having the ability to take actions in case of unexpected events, such as cyber attacks \cite{richter2006towards, chaabanrobustmas}.
To achieve this, an observer/controller architecture is introduced \cite{richter2006towards}. The observer monitors the system and performs tasks such as anomaly detection. The controller receives observations and performs actions in response, such as implementing mitigation strategies.

\gls{CPES} often operate under constraints such as limited access to information and control actions, driven by privacy concerns, regulatory limitations or even requirements for data exchange minimization \cite{icaart25}.
It is essential to consider these constraints when implementing a suitable observer/controller architecture, as access to information impacts the performance of anomaly detection, while limited actions influence the effectiveness of possible controller implementations. 
To consider a robust self-organizing system, a suitable observer/controller architecture must be designed considering access to information, actions. It is essential to find a suitable architecture taking these constraints into account, in order to ensure the functionality and robustness of the existing system, even when disturbances occur.
In this context, previous research has focused on evaluating different observer architectures for anomaly detection, considering limited access to information \cite{icaart25}.
The contribution of this work is to add and develop the controller perspective. An evaluation of different controller actions and architectures is done for an agent-based control use case. The evaluation's outcomes improve existing reaction strategies, thereby enhancing the system's overall robustness in the presence of disturbances or attacks.

The paper is organized as follows. In \autoref{sec:rw}, related work regarding different design variants of the observer/controller architecture for \gls{CPES} is presented. Afterwards, \autoref{sec:obs} discusses relevant architectures for the observer, followed by the presentation of controller actions and architectures in \autoref{sec:cont}. In \autoref{sec:ev}, multiple architecture variants are evaluated, focusing on the controller but taking into account previous observer results. The paper concludes with the discussion in \autoref{sec:diss} and the conclusion in \autoref{sec:conc}.

\section{Related Work}\label{sec:rw}
%In previous work, related work regarding observer architectures for anomaly detection is discussed \cite{icaart25}. 
%When considering observer/controller architectures, the outcomes are quite similar. 
In this section, design variants for the observer/controller architecture are discussed mutually, considering concepts from Organic Computing and for controller architectures for \gls{CPES} in general\footnote{This section extends the related work presented in \cite{icaart25} (focusing on observer architectures) for controller architectures.}. 
\paragraph{Centralized architectures}
In centralized architectures for observer and controller, a single observer and a single controller regulate various components of the entire system and intervene with all entities \cite{tomforde2011observation}. A centralized observer has access to global information knowledge from all control units \cite{tan2020brief, tomforde2011observation, ERHAN202164}, while the controller takes action on the entire system \cite{chen2021cyber}.

\paragraph{Decentralized architectures}
This type of architecture considers one controller and one observer separately for each component \cite{tomforde2011observation}, without any information about other parts of the system \cite{tan2020brief}.
Using a decentralized architecture for the observer and controller, the operation process of the system is decoupled into smaller steps \cite{chen2021cyber}. This might be the case to parallelize steps for efficiency \cite{chen2021cyber}. In this architecture, no communication between the decentralized controllers is assumed \cite{chen2021cyber}.

\paragraph{Distributed architectures}
Distributed architectures are similar to decentralized ones, however, communication between the controllers is possible \cite{chen2021cyber}. Information about local measurements can be communicated among neighboring controllers \cite{tan2020brief}.

\paragraph{Multi-Leveled / hierarchical architectures}
Furthermore, it is also possible to combine the previous architectures into multi-leveled or hierarchical architectures \cite{tomforde2011observation}. In hierarchical architectures, one observer and one controller per component exists, but an additional secondary level of observer and controller is added, responsible for the complete system \cite{tan2020brief, tomforde2011observation}. \\

Even though these architectural models have been theorized and discussed, only very few have been implemented and compared to investigate the effects, particularly under constraints as limited information or control access. Chaaban et al. combine a centralized controller with a decentralized system, presenting a hybrid approach \cite{chaaban2014methodology}. However, access to information or control actions is not discussed. 
In previous work, different anomaly detection architectures are evaluated, considering information availability from the observer perspective only \cite{frost2024detecting, icaart25}. 
This paper thus fills the gap in the joint consideration of architectures for the observer and the controller for application in agent-based \gls{CPES}, taking into account access to information and control measures.
\section{Observer Architectures in \gls{CPES}}\label{sec:obs}
This paper summarizes the outcomes regarding observer architectures for anomaly detection, as presented in previous work \cite{icaart25}. At first, the agent-based control use case is presented, depicting the implementation setup. The entire section refers to the previous work \cite{icaart25}.
\subsection{Use Case Description}
The use case is self-consumption optimization, considering redispatch requests from a grid operator, according to the system concepts presented in \cite{krueger2023redispatch, radtke2023integrating}. The agent system represents a neighborhood grid (energy community), performing the self-consumption optimization. The agents manage \gls{DER} in this neighborhood grid, such as wind turbines, \gls{PV} systems or battery storage systems. Additionally, agents are integrated to control households \cite{icaart25}. 
The setup enables evaluation of how cyber attacks, particularly false data injection attacks, impact system behavior.
Attacks target power values in exchanged messages in the agent system, which mimics realistic threats in \gls{CPES} environments. The details for the scenario and the compromised agent are described in \cite{frost2024}. 
\subsection{Anomaly Detection and Information Availability}
\begin{figure}[ht]
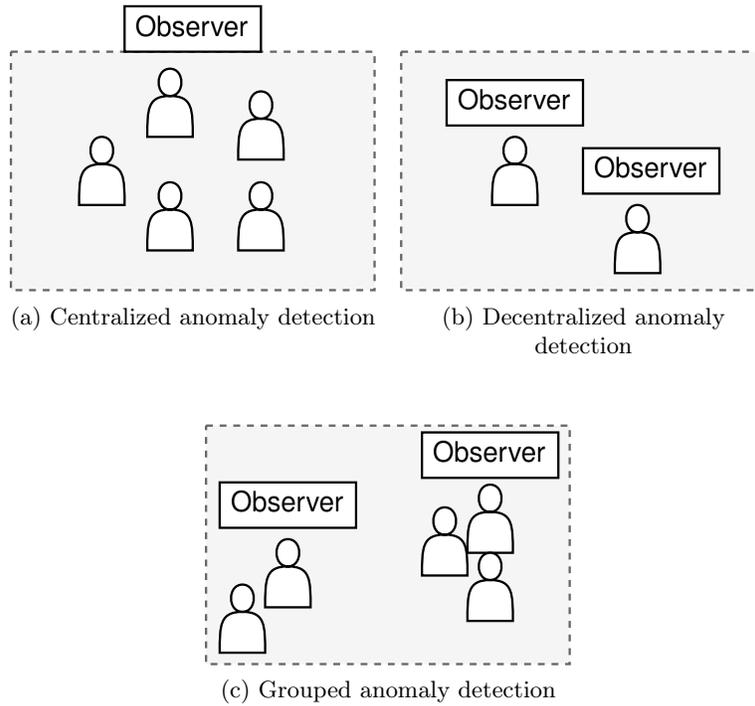

    \captionsetup{farskip=3pt}
    \centering
    \subfloat[\centering Centralized anomaly detection \label{fig:a}]{{\includesvg[width=0.4\textwidth]{zentral.svg}}}\hspace{0.2cm}
    \subfloat[\centering Decentralized anomaly detection\label{fig:b}]{{\includesvg[width=0.4\textwidth]{verteilt.svg}}}\hspace{0.2cm}%\hfill
    \subfloat[\centering Grouped anomaly detection\label{fig:c}]
    {{\includesvg[width=0.4\textwidth]{gruppiert.svg}}}    \caption{Architectures for anomaly detection in agent systems (presented in \cite{icaart25})}
    \label{fig:architectures}
\end{figure}

To detect attacks, an observer is implemented. Various architectures for anomaly detection from the literature were taken into account for the observer implementation. The considered anomaly detection architectures are depicted in \autoref{fig:architectures}.

The centralized anomaly detection in \autoref{fig:a} is implemented for the entire system, considering data of the whole network. In the decentralized anomaly detection architecture (\autoref{fig:b}), one observer for each component of the system exists, only maintaining this component. Therefore, the decentralized observer considers data of only one component. As a third observer architecture, a grouped anomaly detection is presented, as shown in \autoref{fig:c}. This architecture only considers data from a group of components. Two versions of this architecture were implemented. In the first version, a group of a specific unit type is formed. Thus, only the data of this specific group of units is considered (i.e. all wind units). In the second grouped anomaly detection, a random group of \gls{DER} is constructed. 
The following summarizes the implemented architectures and the considered data.

\begin{itemize}
    \item Centralized architecture: data of the whole network is considered.
    \item Decentralized architecture: data of only one component is considered.
    \item Grouped by unit type: data of all units of a certain type is available, e.g., all wind units.
    \item Grouped randomly: data of a random group of \gls{DER} is given.
\end{itemize}

Additionally, the availability of information is discussed for all four architectures. In \gls{CPES}, depending on the application and constraints, some data may not be permitted to be shared, particularly with regard to local \gls{DER} constraints. In a neighborhood grid for self-consumption optimization, limitations regarding the available data for anomaly detection can also be assumed. For this reason, different information levels were defined and considered when performing anomaly detection. The first level only considers the agent and timestamp of the exchanged message (no insight into the message is given). The second level considers communication data in addition to the agent and timestamp, such as message delays and current system traffic. The third level assumes insight into the messages, considering the message content including exchanged values. The final level adds additional data concerning the constraints of the specific \gls{DER} controlled by the agents. The information levels are summarized below.

\begin{enumerate}
    \item Agent and timestamp
    \item Communication data (delays, traffic)
    \item Message content (exchanged values)
    \item Additional data (constraints of the specific \gls{DER})
\end{enumerate}

\subsection{Anomaly Detection Outcomes}
A transformer autoencoder was implemented for anomaly detection due to its strong performance in prior research. The transformer autoencoder was used for all types of architectures and information levels, differing in the available data. Implementation details are described in previous work \cite{icaart25}.

The results of the anomaly detection show that without having the message content available (information levels 1 and 2 only), none of the architectures can reliably detect anomalies. Satisfactory detection is only possible starting from level 3, assuming insight into the exchanged messages. Therefore, detection of anomalies in compromised data is not feasible without having access to the content of the messages.
Of all the architectures, decentralized anomaly detection is the most effective, most likely due to its focus of one agent's behavior. This aligns with the privacy constraints in \gls{CPES}, since local data cannot be assumed to be shared at a central location.
Additionally, the results show that grouping agents does not improve performance and may introduce unnecessary complexity.

\begin{figure}[H]
\begin{center}
 \includesvg[width=0.45\textwidth]{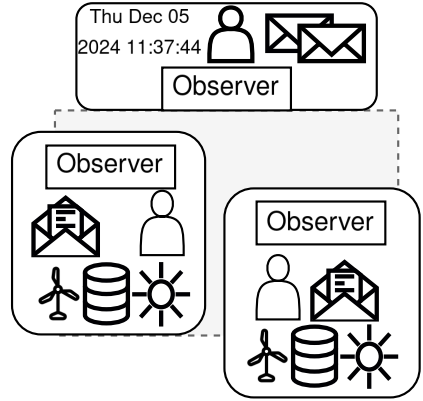}
\caption{Proposed multi-leveled observer architecture (presented in \cite{icaart25})}
 \label{fig:proposed}
\end{center}
\end{figure}

The outcomes furthermore indicate the relevance of multi-leveled architectures. Combining centralized and decentralized detection seems promising in order to detect various kinds of anomalies. In this context, a multi-leveled architecture is proposed, as depicted in \autoref{fig:proposed}. The centralized observer in this architecture operates without message content, to be able to monitor broad system behavior, while the decentralized observer focuses on localized, content-specific anomalies by having access to the message content and considering unit-specific data.
This combination could lead to effective anomaly detection able to detect different kinds of attacks (e.g., both, anomalies in the exchanged values and in the communication behavior). Additionally, it respects privacy constraints, reduces communication load and improves robustness.

\section{Controller Architectures in \gls{CPES}}\label{sec:cont}
Once the observer detects anomalies, it forwards the information to the controller, which is responsible for reacting accordingly. Depending on the actions available to the controller and the components it is allowed to manage, different controller architectures exist. 
In the following, controller architectures for agent-based \gls{CPES} are discussed. First, possible controller actions are derived and discussed. Then, variants of these actions for different architectures are presented. The results of the evaluation of observer architectures described above are taken into account in the discussion in this section.

To examine possible controller architectures in agent-based \gls{CPES}, it is first necessary to collect possible actions the controller can perform. Depending on the scope of these actions and the components affected, these actions are transferred to the respective architectures. A literature review was conducted to determine possible actions that controllers perform in agent-based use cases in \gls{CPES}. The outcomes are presented in \autoref{tab1} and discussed in the following.

\iffalse
% Tabelle ANi
\begin{table}
\centering
\caption{Possible controller actions in agent-based \gls{CPES}}\label{tab1}
\begin{tabular}{l|l}
%\hline
Controller action & Approach\\
\hline
Isolation of compromised assets & \cite{dehghanpour2017survey, zografopoulos2022mitigation, huang2021distributed, kordestani2021observer, liang2022adaptive, zhang2021ndo} \\
Topology adaption due to system state & \cite{hashmicoordinated, dorsch2016intertwined, sharafian2024adaptive, vistbakka2021modelling, bidram2019resilient} \\
Reassignment of tasks & \cite{radtke2023integrating, bidram2019resilient, sahoo2020resilient}\\
%\hline
\end{tabular}
\end{table}
\fi

% Tabelle Template
\begin{table}
\centering
\caption{Possible controller actions in agent-based \gls{CPES}}\label{tab1}
\begin{tabular}{|l|l|}
\hline
Controller action & Approach\\
\hline
Isolation of compromised assets & \cite{dehghanpour2017survey, zografopoulos2022mitigation, huang2021distributed, kordestani2021observer, liang2022adaptive, zhang2021ndo} \\
Topology adaption due to system state & \cite{hashmicoordinated, dorsch2016intertwined, sharafian2024adaptive, vistbakka2021modelling, bidram2019resilient} \\
Reassignment of tasks & \cite{radtke2023integrating, bidram2019resilient, sahoo2020resilient}\\
\hline
\end{tabular}
\end{table}

\paragraph{Isolation of compromised assets}
Once malicious, faulty or compromised assets are detected, a first step is often to isolate and exclude these from the system \cite{dehghanpour2017survey, huang2021distributed}. This way, possible harm is minimized.
To isolate attacked areas or attackers, communication links could be discarded \cite{kordestani2021observer}.  This can include a microgrid formatting during events, using assets which still operate within estimated ranges \cite{xu2021anomaly}. 

\paragraph{Topology adaption due to system state}
Additionally, adaptions of the topology due to the current system state are discussed in the literature. Dorsch et al. present a setting in which agents ask a controller for adaptions as a raise in priority of traffic flows in case of an overload \cite{dorsch2016intertwined}. In other approaches, agents can allocate additional bandwidth if necessary to communicate \cite{hashmicoordinated}. Other works consider the controller breaking down the system into smaller subsystems to regulate complex behaviors after an attack happening \cite{sharafian2024adaptive}.
Additionally, a controller can discard communication links to isolate attackers \cite{kordestani2021observer} and switch off lines \cite{liang2022adaptive}.

\paragraph{Reassignment of tasks}
After excluding a malicious agent from the operation, its tasks can be assigned to another agent in order to ensure robustness of the \gls{MAS}. This approach has also been successfully implemented in other domains, such as warehouse systems \cite{vistbakka2021modelling}.
To do so, the existing available sources are used \cite{bidram2019resilient, sahoo2020resilient}, as grid sources in \gls{CPES} \cite{zografopoulos2023distributed}. \\

\par When comparing different possible actions, it becomes clear that isolating compromised assets is another way to adapt the topology. This method excludes the isolated assets from the topology. Redesigning the topology can also be necessary.

%\subsection{Controller Architectures}\label{ssec:arch}
% The following discusses the architectures described in \autoref{sec:rw} and how these actions can be implemented with limited access.
\subsection{Centralized Controller}
\begin{figure}[H]
\begin{center}
 \includesvg[width=0.4\textwidth]{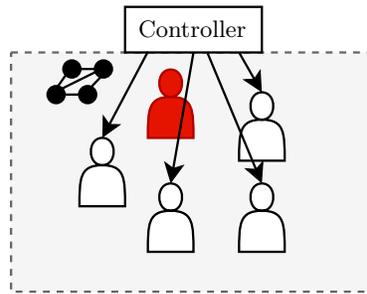}
\caption{Reaction to the malicious agent: the centralized controller sends a new communication topology, excluding the malicious agent from communication.}
 \label{fig:zentralcontrol}
\end{center}
\end{figure}

A centralized controller has access to the entire system and can perform actions that affect all components. Once the observer detects malicious agents, the controller's task is to exclude them from the system in order to minimize possible harm. This can be easily accomplished by informing all other agents to do so, as the centralized controller can directly contact all agents in the system. The process is depicted in \autoref{fig:zentralcontrol}. Additionally, a new communication topology can be calculated for the entire system without the malicious agent.
The centralized controller, having full knowledge of the system, can reassign the compromised agent's task to another agent. For example, if the failing agent was responsible for controlling a \gls{PV} unit, the controller can pass information, data, and tasks to other non-malicious agents to overtake the \gls{PV} control. This ensures that the system's full functionality is maintained while avoiding performance losses.

\subsection{Decentralized/Distributed Controller}
\begin{figure}[H]
\begin{center}
 \includesvg[width=0.4\textwidth]{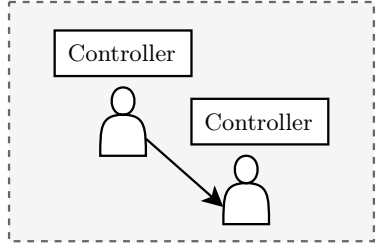}
\caption{Reaction to the malicious agent: the decentralized controller informs its neighbors about the malicious agent.}
 \label{fig:verteiltc}
\end{center}
\end{figure}

The decentralized controller does not have access to the whole system, it is implemented for each component separately. Once a malicious agent is detected, the decentralized controller excludes this agent from the system state. As the controller knows about the agent's neighbors, it can also inform the neighbors directly or the neighboring controllers in case of a distributed control. The neighboring agents can then also exclude the malicious agents from the system state to avoid potential critical situations. The process is shown in \autoref{fig:verteiltc}.
After excluding malicious agents from the system, it is difficult to ensure that the task of the compromised agent is assigned to another one. Due to the decentralized nature of the system and limitations due to privacy etc., the decentralized controller does not always have enough information about other agents to be able to take over this task or transfer it to another agent.
At this point, it would be conceivable for grouped architectures to exist. A group of decentralized controllers that are responsible for a system type (similar to the grouped observers per system type in \cite{icaart25}) would have information about the control of that system, since it functions similarly to the control of their own agent's system. Furthermore, task catalogs could be integrated that are accessed as soon as a task is no longer fulfilled.
Nevertheless, the exclusion of faulty agents already represents a significant step towards the robustness of the system, as this ensures that compromised agents cannot cause widespread damage.

\subsection{Multi-leveled Controller}
\begin{figure}[ht]
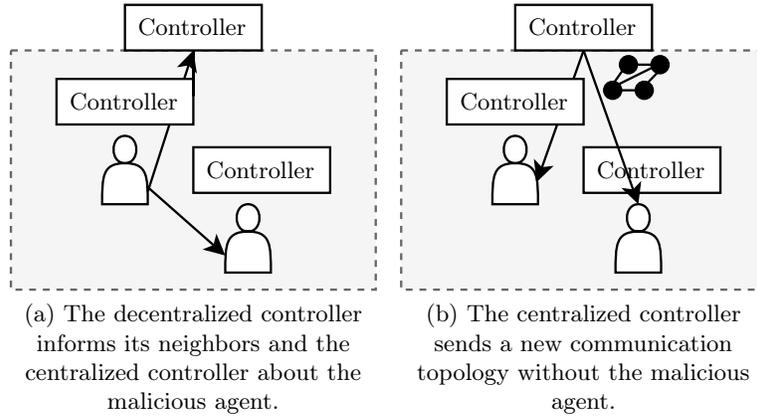

    \captionsetup{farskip=3pt}
    \centering
    \subfloat[\centering The decentralized controller informs its neighbors and the centralized controller about the malicious agent.\label{fig:5a}]{{\includesvg[width=0.4\textwidth]{multi_level.svg}}}\hspace{0.2cm}
    \subfloat[\centering The centralized controller sends a new communication topology without the malicious agent.\label{fig:5b}]{{\includesvg[width=0.4\textwidth]{multi_level_2.svg}}}\hspace{0.2cm}%\hfill
 \caption{Reaction to the malicious agent: multi-leveled controller.}
    \label{fig:ml}
\end{figure}

In a multi-leveled architecture, additionally to the decentralized controller, a secondary instance is added: the centralized controller. In this way, control actions can stem from different directions, affecting different agents. If the centralized observer detects a compromised agent, it can directly create a new communication topology without it and send it to the system. The case in which the decentralized observer/controller detects a faulty agent is shown in \autoref{fig:ml}. Here, this controller first informs its neighbors about the compromised agent, as in the case of the exclusively decentralized controller. It can also inform the centralized controller about its observations, which can then create a new communication topology without the faulty agent and send it to the system.

Afterwards, the task can be assigned to other agents by the centralized controller. In this way, it is ensured that detected anomalies are taken into account in both architectures (decentralized and centralized). A multi-level observer architecture would therefore be well covered here.

\section{Evaluation: Designing a Controller for Agent-Based Control}\label{sec:ev}
This section evaluates the previously presented controller architectures. The agent-based system presented for the anomaly detection is used \cite{icaart25}. The cyber attacks added to the system were taken into account in accordance with the preliminary work and a controller was added next to the observer. 
The results of the observer evaluation are taken into account for the implementation. A multi-leveled observer is assumed, which combines a centralized observer with access to data of the entire system and a decentralized observer with access to local constraints of individual components. 
Various controller architectures are now evaluated for this observer.
The metrics presented in the preliminary work are used to evaluate the impact of the different controller architectures \cite{frost2024}. The duration of an optimization (convergence speed), the quality of the solution and the number of messages exchanged are examined, as these metrics provide detailed insights into the systems properties. Previous work has also looked at the average duration of the messages exchanged, but as compromised data does not affect this, this metric is excluded here.

To evaluate the impact of different controller architectures, three phases of operation were investigated, as depicted in the following.
\begin{enumerate}
    \item Normal operation: no attack or disturbances occurring.
    \item Disruption: the false data injection attack is present, introducing a compromised agent to the system.
    \item Control active: the controller reacts to the detected malicious agent. 
\end{enumerate}

The event of phase 2 (disruption) starting is depicted as incident in the figures, starting from time step 20. The controller starts to get active at time interval 36. 
For the ranges in order to differ between normal operation and disruption, the margins of the operation were collected using gathered data from normal operation. The upper and lower acceptable margins are maximum and minimum values from the normal operation and the target value is determined using the mean value of normal operation, as described in \cite{frost2024}. This helps to directly detect deviation from normal behavior.

\subsection{Centralized Controller}
Once a malicious agent is detected, the centralized controller determines a new communication topology for the agent system (without the malicious one). All agents are informed of this change and take over the new topology.
The convergence speed, the solution quality and the exchanged messages for all three phases for a centralized control are discussed in the following. 

\begin{figure}[H]
\begin{center}
 \includesvg[width=\textwidth]{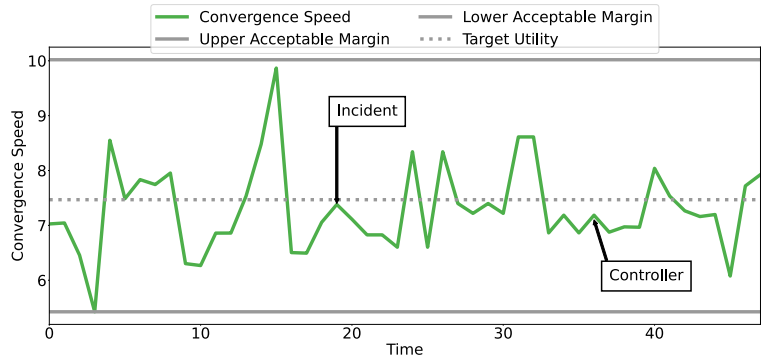}
\caption{Centralized controller: convergence speed (optimization duration) for normal operation, disruption and control phases.}
 \label{fig:centralnegotiation}
\end{center}
\end{figure}

Regarding the optimization duration, shown in \autoref{fig:centralnegotiation}, it is visible that no deviations from the ranges exist. The convergence speed is always within the determined normal operation ranges, for the normal operation, the incident occurring with and without controller being active.

\begin{figure}[H]
\begin{center}
 \includesvg[width=\textwidth]{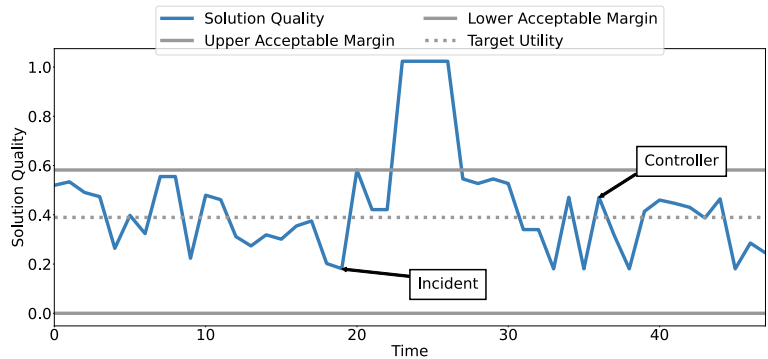}
\caption{Centralized controller: solution quality (performance) for normal operation, disruption and control phases.}
 \label{fig:centralsolution}
\end{center}
\end{figure}

For the solution quality, depicted in \autoref{fig:centralsolution}, while the disruption is active and the controller is not yet active, the quality is outside the permitted ranges.

As the agents solve a minimization problem (the objective is to minimize the difference of the determined active power to the optimal self-consumption rate), high, wide outliers are interpreted negatively. It can therefore be seen that the attack has an effect on the solution quality.
Once the controller becomes active, the solution quality is within the normal, acceptable limits. The controller's action therefore contributes to the agent's solution moving back into permitted, valid ranges.

\begin{figure}[H]
\begin{center}
 \includesvg[width=\textwidth]{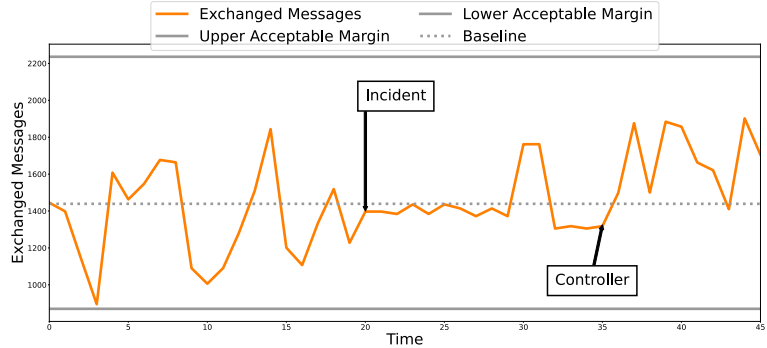}
\caption{Centralized controller: number of exchanged messages for normal operation, disruption and control phases.}
 \label{fig:centraltraffic}
\end{center}
\end{figure}

The number of messages exchanged is shown in \autoref{fig:centraltraffic} for the centralized controller. It can be seen that these results do not show any special features in all three phases. The average number of messages exchanged is within the permissible ranges (within lower and upper acceptable margin from normal operation), with no noticeable outliers.

\subsection{Decentralized Controller}
Once the observer detects a malicious agent, the decentralized controller excludes this agent from its own system state as a first step. It will not send anymore messages to this agent and exclude it from existing states. Afterwards, the decentralized controller informs its neighbors about the malicious agent, which will then also exclude it from their own system state. The other agents will additionally inform their own neighbors about the malicious one. In this way, the information about the compromised agent is forwarded over the whole system.
The convergence speed, solution quality and number of exchanged messages for normal operation, the disruptions and considering only the decentralized controller, are presented in the following.

\begin{figure}[H]
\begin{center}
 \includesvg[width=\textwidth]{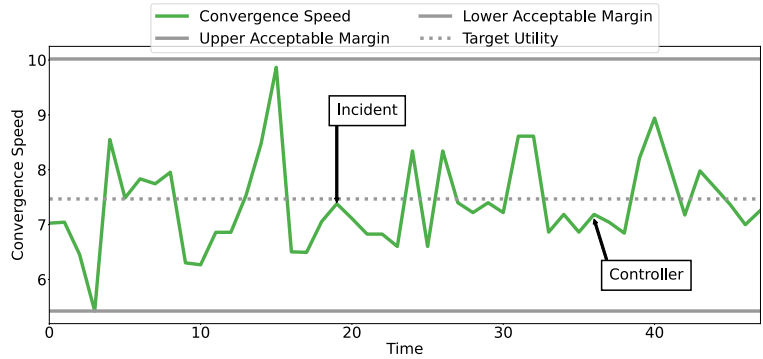}
\caption{Decentralized controller: convergence speed (optimization duration) for normal operation, disruption and control phases.}
 \label{fig:decentralnegotiation}
\end{center}
\end{figure}

The results for the decentralized for the convergence speed (optimization duration), are shown in \autoref{fig:decentralnegotiation}.
The decentralized controller also shows no particularities in the length of optimizations, in any of the three phases (normal, disruption, control).

\begin{figure}%[H]
\begin{center}
 \includesvg[width=\textwidth]{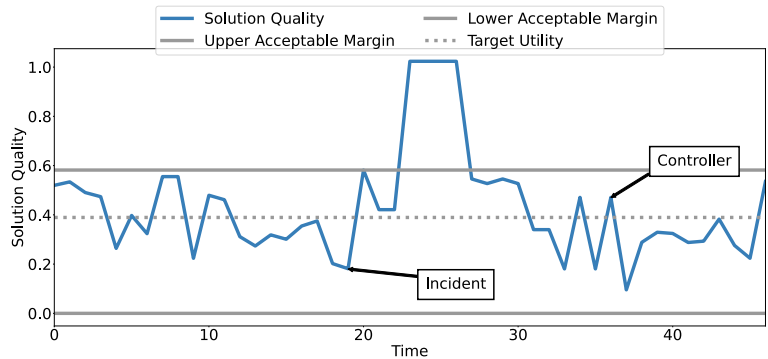}
\caption{Decentralized controller: solution quality (performance) for normal operation, disruption and control phases.} 
 \label{fig:decentralsolution}
\end{center}
\end{figure}

The results for the decentralized controller for the solution quality are depicted in \autoref{fig:decentralsolution}. The behavior is similar to that of the centralized controller. As soon as the decentralized controller becomes active and implements his measures, the solution quality is within the acceptable ranges. Here, too, it can be seen that the decentralized controller manages to move the agents' solution into a permissible, good range through its measures.

\begin{figure}%[H]
\begin{center}
 \includesvg[width=\textwidth]{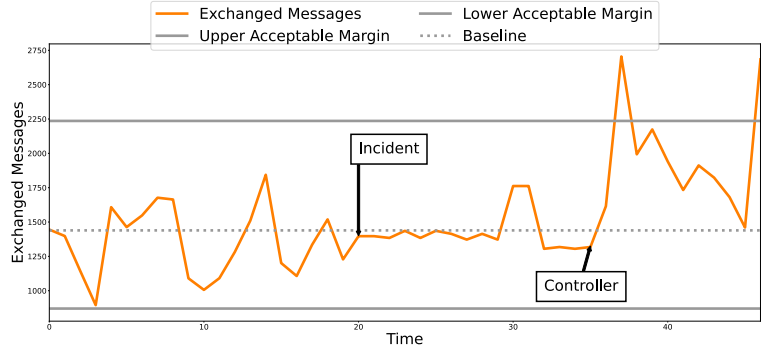}
\caption{Decentralized controller: number of exchanged messages for normal operation, disruption and control phases.}
 \label{fig:decentraltraffic}
\end{center}
\end{figure}

With regard to the number of messages exchanged with a decentralized controller, shown in \autoref{fig:decentraltraffic}, there is a noticeable difference to the centralized controller. After the controller becomes active, there are isolated outliers that are above the limits defined as normal behavior by the normal operation. This can be explained by the fact that the decentralized controller reports agents detected as malicious to its neighbors, which then also forward this information to their own neighbors. In contrast to the centralized controller, which sends the new topology directly to the individual agents, many messages are required until all agents are informed.

\section{Discussion}\label{sec:diss}
The results of the different controller architectures contain relevant findings to be considered in the design of observer/controller architectures.
It was shown that both controller architectures ensure that a sufficient performance is reached. The fact that the malicious agent is excluded from the system by both controllers means that this agent is no longer part of the solution and can no longer influence it. The solutions are therefore valid and feasible again. In addition, the solutions are also back within the range of normal operation, which suggests acceptable performance. This metric therefore shows the effectiveness of both controllers.

However, there are also recognizable differences in the controller architectures. Although both architectures have a positive effect on the solution quality and there are no significant differences in the duration of the respective optimizations, there are differences in the number of messages exchanged.
Using the decentralized controller, a large number of messages is exchanged as soon as faulty agents are detected and the controller becomes active. In contrast to the centralized controller, the number of messages is even outside the limits that were determined based on the normal operation.
This shows the differences between the architectures. While the centralized controller can inform all agents directly about the malicious one, this is not possible for the decentralized controller. It only knows the agent's neighbors and can thus only inform these. The decentralized controller is therefore dependent on other decentralized controllers to disseminate the message.
Accordingly, the number of messages exchanged increases.
The interpretation of the fact that the number of messages is outside the limits of the normal operation depends very much on the application in the energy system and the requirements of the agent system. 
The controllers have initially both ensured that the faulty agents are excluded from the system. This means that the solution is permissible and the system is functioning. With decentralized control, there is no other option to inform other agents (or their controllers). This would require further information, which can not be assumed to be available in such a decentralized setting.
Therefore, depending on the application, it can be discarded if the number of messages exchanged increases significantly, as is it no problem for the system's functionality.
An alternative solution must be sought for use cases in which the number of exchanged messages is to be kept low due to bandwidth limitations.

The results clearly demonstrate that different controller architectures have an influence on the system performance. The controller to be implemented must be evaluated depending on the available measures.
These evaluation results support the statement that different observer/controller architectures have different effects on robustness. The choice of architecture must therefore not only be made according to the complexity, heterogeneity and size of the system, as described in \cite{tomforde2011observation}, but also depending on the availability of information \cite{icaart25} and the existing actions, as shown in this evaluation.
The best-suited controller architecture depends heavily on the use case and the requirements placed on it. It is important to consider which metrics and requirements are used to evaluate the system. This work provides an initial foundation on which to build when designing observer/controller architectures for agent-based control use cases in energy systems.

\section{Conclusion}\label{sec:conc}
In agent-based \gls{CPES}, there are constraints such as privacy, regulatory restrictions or other requirements that prevent local \gls{DER} data from being available at a central location.
To ensure the robustness of an agent-based system, controllers and observers can be added to monitor the system and intervene in the event of disturbances. 
However, when implementing these, the constraints and general conditions of the \gls{CPES} must also be taken into account. This also covers access to information and to control actions.
Previous work has shown that different architectures and available information have a significant impact on the performance of the observer's anomaly detection \cite{icaart25}. This work extends the previous work to include the perspective of the controller. Various possible architectures were presented for the controller, which evaluate possible measures based on controlled components.
We could demonstrate that controller architectures have an influence on the system performance and therefore must be considered in the design of robust controlled self-organization.

Examining the available information and actions is essential for identifying a suitable observer/controller architecture.
Thus, this work establishes a foundation for enhancing the robustness of agent-based control systems by presenting outcomes that optimize the design of controlled self-organization.
Future research should involve an investigation using a specific application going beyond the one given here. In doing so, the requirements for this application must be determined in detail. This allows an evaluation of which observer/controller architecture best fulfills the specific requirements for a real-world application.
In general, setting up a design process to identify the best fitting observer/controller architecture for specific use cases including a tool chain to support this process would be of high relevance in this field.
%Furthermore, additional architectures could be implemented. The work discusses possible controller architectures and also includes initial ideas for the development of multi-leveled controller architectures.

\section*{\uppercase{Acknowledgements}}
The authors would like to thank the German Federal Government,
the German State Governments, and the Joint Science Conference
(GWK) for their funding and support as part of the NFDI4Energy
consortium. The work was partially funded by the German Research Foundation (DFG) – 501865131 within the German National
Research Data Infrastructure (NFDI, www.nfdi.de).
\bibliographystyle{splncs04}
\bibliography{bibliography}
\end{document}